\documentclass[aps,prl,twocolumn,superscriptaddress]{revtex4-1}
\usepackage{amsmath}
\usepackage{amssymb}
\usepackage{graphicx}
\usepackage{bm}

\begin{document}

\title{Thermal and electrical quantum Hall effects in ferromagnet--topological insulator--ferromagnet junction}

\affiliation{ 1. Institut f\"ur Theoretische Physik, Universit\"at Hamburg,
Jungiusstr 9, D-20355 Hamburg, Germany}
\affiliation{
Shamoon College of Engineering, 
56 Bialik St.  P.O.Box 950, Beer-Sheva 84105, 
Israel}
\affiliation{
Max-Planck-Institut f\"ur Physik komplexer Systeme
N\"othnitzer Str. 38,  01187 Dresden, Germany}

\author{A. L. Chudnovskiy}
\affiliation{ 1. Institut f\"ur Theoretische Physik, Universit\"at Hamburg,
Jungiusstr 9, D-20355 Hamburg, Germany}
\affiliation{
Max-Planck-Institut f\"ur Physik komplexer Systeme
N\"othnitzer Str. 38,  01187 Dresden, Germany}

\author{V. Kagalovsky }
\affiliation{
Shamoon College of Engineering, 
56 Bialik St.  P.O.Box 950, Beer-Sheva 84105, 
Israel}
\affiliation{
Max-Planck-Institut f\"ur Physik komplexer Systeme
N\"othnitzer Str. 38,  01187 Dresden, Germany}

\date{\today}

\begin{abstract} 
We present the theoretical description for a class of experimental setups that measure quantum Hall coefficients in ferromagnet-topological insulator-ferromagnet (FM-TI-FM) junctions. We predict that varying the magnetization direction in ferromagnets, one can change the induced Hall voltage and transverse temperature gradient from the maximal values, corresponding to the quantized Hall coefficients, down to their complete suppression to zero. We provide detailed analysis of thermal and electrical Hall resistances as functions of the magnetization directions in ferromagnets, the spin-scattering time in TI, and geometrical positions of FM leads and measurement contacts.  
\end{abstract}


\maketitle

Topologically protected helical edge states represent the trademark of topological insulators (TI)
 \cite{Kane-Mele,Bernewig-Zhang,Koenig07,Hasan-Kane10}. Those states provide ideal channels of spin-polarized electric currents, inspiring the use of TI as elements of spintronic devices. A variety of magneto-transport effects in three dimensional TI coupled to FM in layered structures has been studied in a series of recent papers, in particular, the effects of the exchange coupling between FM and TI \cite{Yokoyama2011,Yokoyama-heat}, the coupled spin and charge diffusion \cite{original}, and the spin-torque effects  \cite{46}.   

\begin{figure}
\includegraphics[width=8cm]{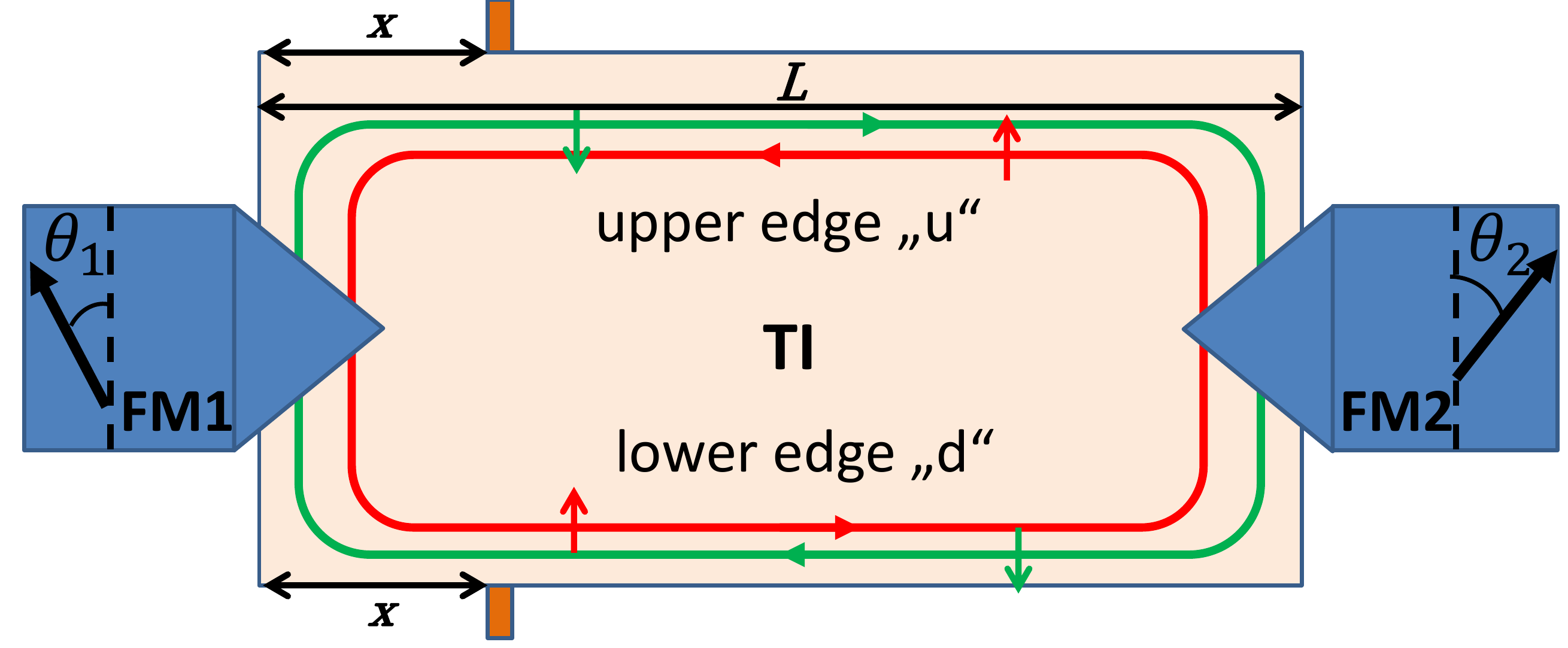} %
\vskip -0.1cm
\caption{(Color online) Proposed experimental setup of FM-TI-FM junction. In contrast to layered junctions, the ferromagnets are connected to TI by point contacts.  Spin-$\uparrow$ and spin-$\downarrow$ edge states have opposite chirality. Measurements of the transverse voltage and temperature gradients are performed at equal distances $x$ from the left contact on both edges. 
\label{fig-Setup}}
\end{figure}

In this paper, we explore the  strong similarity of transport phenomena  in TI and in the quantum Hall systems provided by the existence  of topologically protected conducting edge states along with the  insulating bulk \cite{Kane-Mele,Bernewig-Zhang}. In the presence of time-reversal symmetry, proper to TI, the quantum Hall effect is hidden from experimental observation though. For instance, each spin-resolved edge state would contribute a unit quantum of the Hall conductance although no external magnetic field is applied. However, in contrast to the quantum Hall systems, the chiral edge states with opposite spin-projections propagate in the opposite directions, which results in the exact cancellation of contributions to the total Hall resistance in TI from the two counter-propagating edge states \cite{Bernewig-Zhang}. The quantum Hall effect can be revealed, if the symmetry between the edge states propagating along the opposite edges of the sample is broken, for example, by contacting TI to ferromagnets \cite{Bernewig-Zhang,Kane-Mele}. In particular, we consider an experimental setup, which consists of a two-dimensional TI positioned between the two FM  metals in a FM-TI-FM point contact junction, as it is shown schematically in Fig. \ref{fig-Setup}. The ferromagnetic electrodes provide the way of spin-selective injection of electrons in TI. So, for a completely polarized ferromagnet it is possible to contact a single chiral spin-polarized edge state, which results in quantized values of electrical and thermal Hall conductances proper to the lowest Landau level of the integer quantum Hall effect  $G_{\mathrm{Q}}=dI/d V_H=e^2/h$,   $K_Q=dQ/d T_{\perp}=(\pi^2 k_B^2/3 h)T$. 

The ideal picture of electric and thermal quantum Hall effects can be spoiled by magnetic impurities that are present in TI, and introduce the back-scattering between the counter-propagating edge states. Although high concentrations of magnetic impurities can lead to the Anderson localization of the edge states \cite{Altshuler2013}, the chiral nature of edge states remains intact at moderate concentrations of magnetic impurities, and the effect of scattering by magnetic impurities can be taken into account by introducing the spin-scattering time \cite{Tanaka2011}. 

\begin{figure}
\includegraphics[width=8cm]{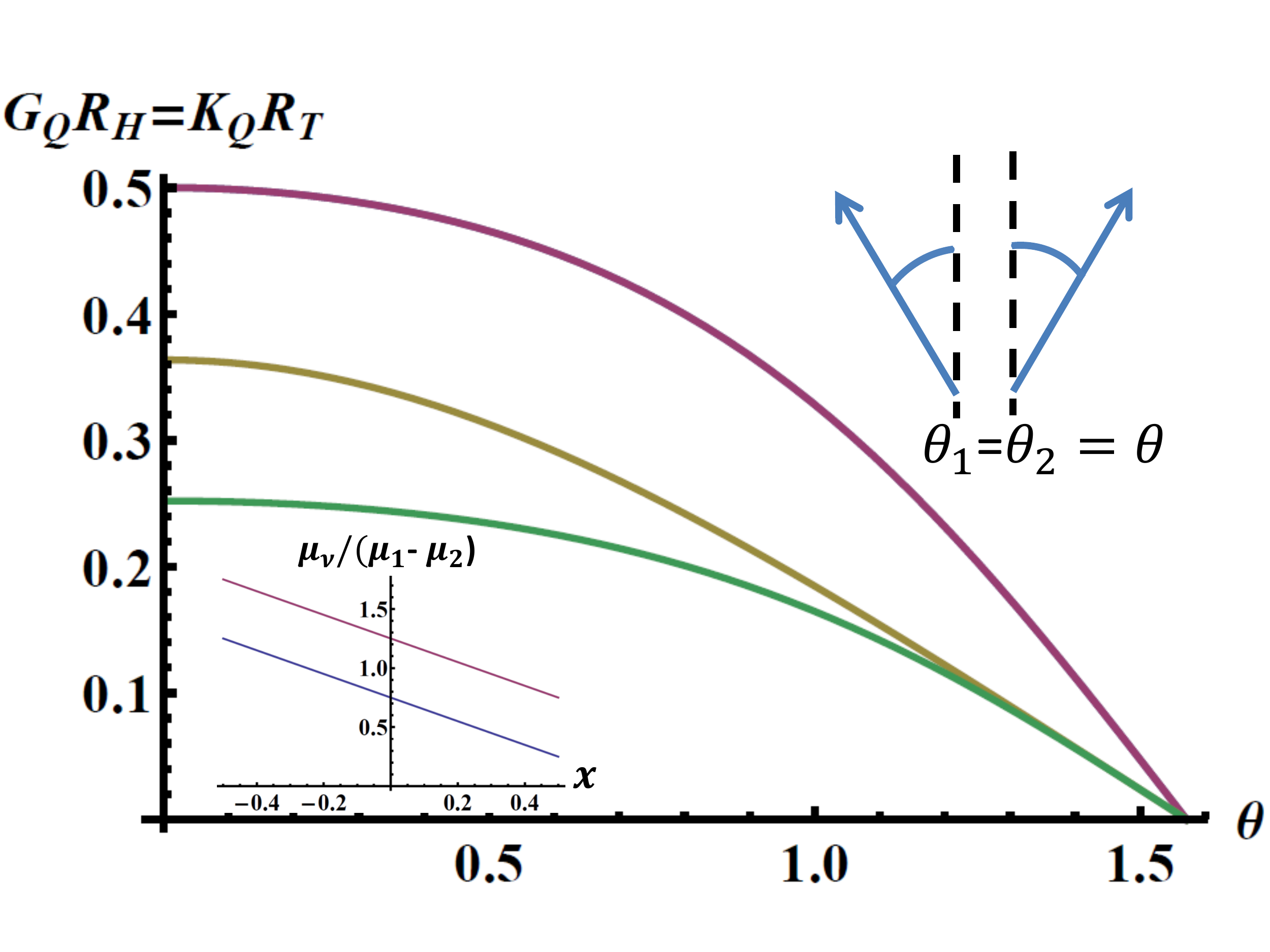}%
\vskip -0.4cm
\caption{(Color online) Dimensionless electrical and thermal Hall resistances as function of the angle $\theta$. The upper curve 
corresponds to symmetric configuration with the  polarizations of contacts $p_1=p_2=1$. This curve is independent of the spin-scattering strength. The two lower curves correspond the setup with one of the leads being a nonmagnetic metal, while the other one is an ideal ferromagnet: $p_1=1$, $p_2=0$.  In that case, the Hall coefficients are suppressed by spin-scattering: the spin-scattering strength is $\xi=0.1$ for the  mid curve, and  $\xi=10$ for the lowest curve.   The total dimensionless conductances of the contacts  $g_{1}=g_{2}=1$ for all curves.  Insets show the magnetization directions in ferromagnets (arrows) and the direction of the spin quantization axis in TI (dashed line) adopted for all three curves, and  the effective chemical potentials at the upper $\mu_{\mathrm u}$ and the lower $\mu_{\mathrm d}$ edges  as functions of coordinate along the edge.  The length of the sample $L=1$.
\label{fig-RH_theta}}
\end{figure}

In the generic case, the point contact between TI and FM  is characterized by a spin-dependent dimensionless conductance $g_{\sigma}$, where $\sigma=\uparrow,  \downarrow$ denotes the spin-projection of the edge state. The total dimensionless conductance of the contact is given by the sum $g=g_{\uparrow}+g_{\downarrow}$. Due to the tunneling magnetoresistance effect \cite{Slonczewski-Sun2007}, the contact conductance depends on the angle $\theta$ between the magnetization in the ferromagnet and the spin-quantization  axis in the TI  
\begin{equation}
g_{\sigma}= g (1\pm p \cos\theta)/2, 
\label{Gsigma}
\end{equation}
where $p$ denotes the contact polarization of FM (for the detailed derivation see Refs. \citep{Slonczewski-Sun2007,JMMM}). 
The angular dependence of the contact magnetoresistance opens the possibility to control the induced Hall voltage and transverse temperature gradient by changing the magnetization direction in FM leads. In what follows we present a detailed calculation of quantum Hall coefficients for different magnetization directions, positions of the FM leads and measurement contacts, and spin-scattering strengths. We find that the  dimensionless electrical and thermal Hall resistances turn out to be equal to each other,  
\begin{equation}
G_Q R_H=K_Q R_{T}=\mathcal{F}, 
\label{F}
\end{equation}
where the factor $\mathcal{F}$ depends on contact conductances, polarizations and magnetization directions in ferromagnets, and the spin-scattering time. In the case of identical ferromagnets with equal angle $\theta$ between the magnetization and spin-quantization axis of TI, the factor $\mathcal{F}$ turns out to be remarkably independent of the scattering time. Its analytical expression reads   
\begin{equation}
 \mathcal{F}(g,  p, \theta)=\frac{2p\cos\theta}{4-g+gp^2\cos^2\theta}, 
\label{Fsym}
\end{equation}
where $g$ denotes the total dimensionless conductance of each contact (see the upper curve in Fig. \ref{fig-RH_theta}). It follows from Eq. (\ref{Fsym}) that the Hall coefficients retain their quantized values in the case of completely polarized ferromagnets ($p=1$) with  magnetizations parallel to the spin-quantization axis in the TI ($\theta=0$). 
In the opposite case when electrons injected in the TI are completely unpolarized ($p=0$) or, equivalently, the magnetization of FM electrodes is perpendicular to the spin-quantization axis in the TI ($\theta=\pi/2$), the factor $\mathcal{F}$ in Eq. (\ref{Fsym}) equals to zero, indicating the vanishing Hall voltage and transverse temperature gradient. The charge and thermal quantum Hall effects disappear in complete agreement with the situation in quantum spin Hall system \cite{Kane-Mele,Bernewig-Zhang}. Eqs. (\ref{F}), (\ref{Fsym}) constitute the main result of this paper. 

To get a qualitative understanding of  the role of the ferromagnetic contacts to reveal the charge and thermal (Leduc-Righi effect \cite{LL-Kinetics}) quantum Hall effects in FM-TI-FM junction, let us consider the propagation of spin-up electrons through the setup shown in Fig. \ref{fig-Setup}.  Due to the chirality of the edge states, the spin-up electrons from the left contact propagate along the lower edge. Their distribution has the chemical potential and the temperature of the left FM lead. Analogously, the spin-up electrons from the right contact propagate along the upper edge, carrying the chemical potential and the temperature of the right FM lead.  Thus, in presence of finite longitudinal electrical or heat current between the FM electrodes, a chemical potential difference (voltage) between the edges is created that is perpendicular to the electrical current. At the same time, there is a counter-propagating spin-down edge state in TI. For that state the Hall effect has the opposite sign. If the leads are spin-unpolarized, the Hall voltages created by the spin-up and spin-down edge states compensate each other exactly resulting in zero net effect. For the ferromagnetic leads, however, the contact conductances for spin-up and spin-down electrons differ, the compensation of contributions from spin-up and spin-down edge states does not take place any more, resulting in the finite Hall voltage and transverse temperature gradient. 

Let us now present a general description of FM-TI-FM junction in terms of kinetic equations for distribution functions of the edge states. Electrons in each edge state are described by a distribution function $f_{\nu\sigma}(x,t; \epsilon)$, where $\nu$ denotes the position of the  edge ($\nu=\mathrm{u, d}$),  $\sigma=\uparrow, \downarrow$ is a spin-index, which also determines the chirality of the edge state, and $x$ is the coordinate along the edge. Since we consider the elastic scattering only, the energy $\epsilon$ is conserved.  The  equations describing the evolution of the distribution functions read  
\begin{eqnarray}
\nonumber 
(\partial_t-v\partial_x)f_{\mathrm{u}\uparrow}=-\frac{1}{\tau}(f_{\mathrm{u}\uparrow}-f_{\mathrm{u}\downarrow}) , \\
\nonumber 
(\partial_t+v\partial_x)f_{\mathrm{u}\downarrow}=-\frac{1}{\tau}(f_{\mathrm{u}\downarrow}-f_{\mathrm{u}\uparrow}) , \\
\nonumber 
(\partial_t+v\partial_x)f_{\mathrm{d}\uparrow}=-\frac{1}{\tau}(f_{\mathrm{d}\uparrow}-f_{\mathrm{d}\downarrow}) , \\
(\partial_t-v\partial_x)f_{\mathrm{d}\downarrow}=-\frac{1}{\tau}(f_{\mathrm{d}\downarrow}-f_{\mathrm{d}\uparrow}) .
\label{KinEq}
\end{eqnarray}
Here $v$ denotes the Fermi velocity of the edge states, which has opposite signs for the counterpropagating states. 
The spin-scattering by magnetic impurities is taken into account phenomenologically, introducing the  spin-scattering time $\tau$ \cite{validity_t0}. The influence of ferromagnets is described by imposing the boundary conditions on the distribution functions at the positions of the TI-FM contacts. Let $L$ be the length of the TI sample, and the contacts are positioned at   $x=\pm L/2$  (see Fig. \ref{fig-Setup}) . 
 
\begin{figure}
\vskip -.5cm
\includegraphics[width=8cm]{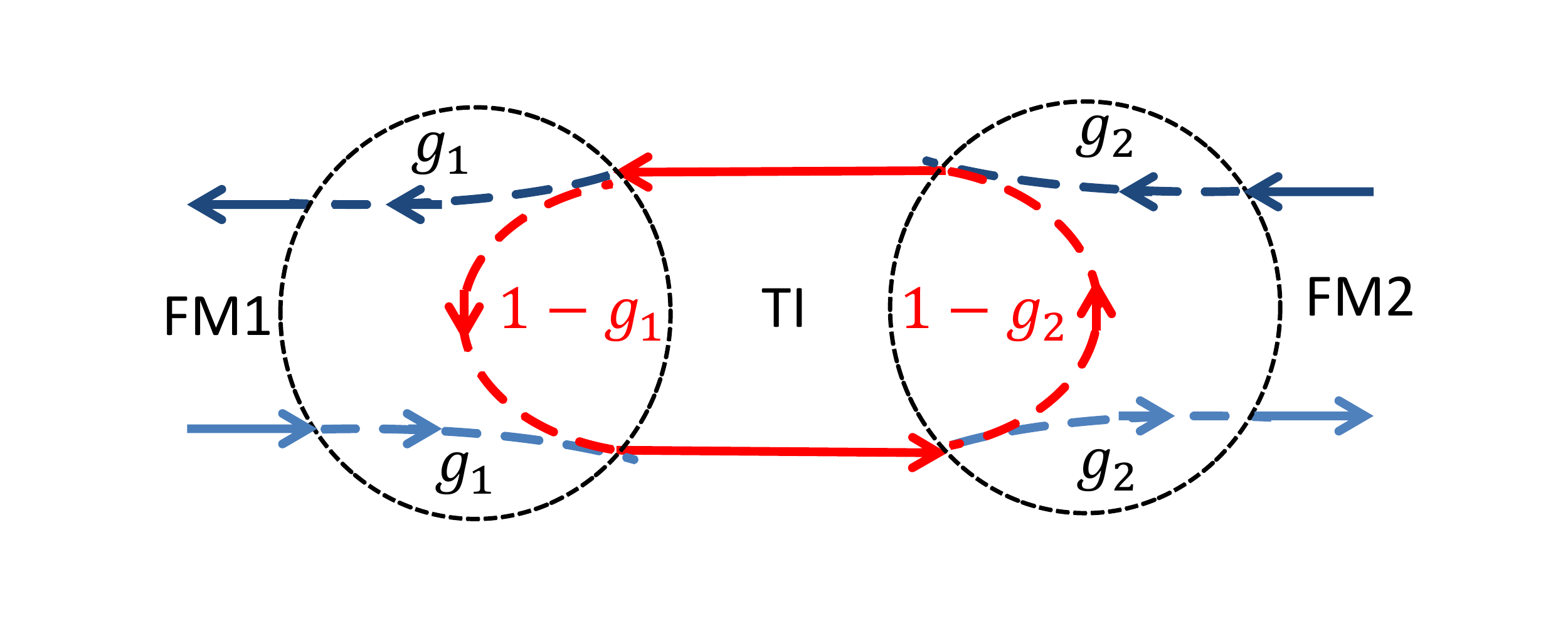}%
\vskip -.5cm
\caption{(Color online) Scheme of scattering in FM-TI-FM junction for a single chiral spin-polarized edge state.    
\label{fig-Scatt}}
\end{figure}
We describe the contacts between ferromagnets and TI using Landauer-B\"uttiker scattering matrix formalism \cite{Buettiker}.
Consider a single spin channel in more detail (see Fig. \ref{fig-Scatt}).  We assume that the phase coherence is lost on the length which is much shorter than the length of the edge channel.  In that case the contacts between the edge state and external reservoirs should be described in terms of transmission probabilities. For example, an electron  coming from the upper edge to the contact 1 is absorbed into the lead FM$1$ with the probability $g_{1\sigma}$, which equals to the dimensionless spin-dependent contact conductance, and it is reflected from the contact into the lower edge with the probability $1-g_{1\sigma}$. At the same time, the incoming wave from  FM$1$ goes to the lower edge with the probability $g_{1\sigma}$. Analogous relations determine the scattering at the contact 2. The probability conservation at the contact 1 ($x=-L/2$) and 2 ($x=L/2$) imposes the boundary conditions to Eqs. (\ref{KinEq}), which read 
\begin{eqnarray}
\nonumber 
f_{\mathrm{d}\uparrow}(-L/2, t) = g_{1\uparrow} F_1+(1-g_{1\uparrow})f_{\mathrm{u}\uparrow}(-L/2, t), \\
\nonumber
f_{\mathrm{u}\downarrow}(-L/2, t) = g_{1\downarrow} F_1+(1-g_{1\downarrow})f_{\mathrm{d}\downarrow}(-L/2, t), \\
\nonumber 
f_{\mathrm{u}\uparrow}(L/2, t) = g_{2\uparrow} F_2+(1-g_{2\uparrow})f_{\mathrm{d}\uparrow}(L/2, t), \\
f_{\mathrm{d}\downarrow}(L/2, t) = g_{2\downarrow} F_2+(1-g_{2\downarrow})f_{\mathrm{u}\downarrow}(L/2, t), 
\label{BC}
\end{eqnarray}
where  $F_1$ and $F_2$ denote the Fermi-Dirac distributions in the ferromagnets. 

Stationary solutions of  Eqs. (\ref{KinEq}) with boundary conditions Eqs. (\ref{BC}) contain the full information about the nonequilibrium distribution functions of the edge states, which are  represented as linear combinations of the distribution functions $F_1(\epsilon)$, $F_2(\epsilon)$ in ferromagnets with coefficients depending on partial conductances  of the contacts $g_{\nu\sigma}$ and the relaxation time $\tau$ 
\begin{equation}
f_{\nu\sigma}(x)=\alpha_{\nu\sigma}(x) F_1+\beta_{\nu\sigma}(x) F_2,  
\label{sol}
\end{equation}
where the particle conservation insures $\alpha_{\nu\sigma}(x)+\beta_{\nu\sigma}(x)=1$, and the coefficients $\alpha$ and $\beta$ are linear functions of $x$.  
The temperature difference or the voltage difference between the left ($F_1$) and right ($F_2$) reservoirs results in  non-equilibrium  distribution functions on the upper and the lower edges. Still one can determine an effective temperature $T_{\nu}$ and an effective chemical potential $\mu_{\nu}$ at the edge $\nu$ by coupling the edge to a system in thermodynamical equilibrium with the Fermi-Dirac distribution function $F_{\nu}(\epsilon)$, which we refer to as the thermometer.  The temperature and the chemical potential of the Fermi-Dirac distribution, at which there is no net heat and particle flow between the thermometer and the edge can be defined as the effective temperature and the effective chemical potential \cite{Sivan-Imry,Pierre,Nazarov-Blanter}. Mathematically, we obtain      
\begin{eqnarray}
\nonumber 
\frac{1}{2} \int (f_{\nu\uparrow}(\epsilon)+f_{\nu\downarrow}(\epsilon)) d\epsilon= \int \left[\exp\left(\frac{\epsilon-\mu_{\nu}}{k_B T_{\nu}}\right)+1\right]^{-1} d\epsilon , \\
 \label{particles}\\ 
 \nonumber 
\frac{1}{2}\int (f_{\nu\uparrow}(\epsilon)+f_{\nu\downarrow}(\epsilon)) \epsilon d\epsilon= \int \epsilon\left[\exp\left(\frac{\epsilon-\mu_{\nu}}{k_B T_{\nu}}\right)+1\right]^{-1}  d\epsilon, \\  
\label{energy} 
\end{eqnarray} 
where  we assumed that the density of states in the edge channel does not depend on energy. Substituting  the distribution functions Eq. (\ref{sol}) into Eqs. (\ref{particles}), (\ref{energy}), we obtain for the effective chemical potentials and effective temperatures at the edges the following relations 
\begin{eqnarray}
\mu_{\nu}(x)=\frac{1}{2} \sum_{\sigma=\uparrow, \downarrow} \alpha_{\nu\sigma}(x) \mu_1+\beta_{\nu\sigma}(x) \mu_2, 
\label{mu_eff} \\
T^2_{\nu}(x)=\frac{1}{2} \sum_{\sigma=\uparrow, \downarrow} \alpha_{\nu\sigma}(x) T^2_1+\beta_{\nu\sigma}(x) T^2_2. 
\label{T_eff}
\end{eqnarray}
The effective chemical potentials as well as effective temperatures squared change linearly with the coordinate $x$ along the edge (see Inset to  Fig. \ref{fig-RH_theta}). Equal slopes of both curves provide for the Hall voltage independent of the position of the measurement contacts. Similar dependence is observed for the effective temperatures squared. 
\begin{figure}
\includegraphics[width=8cm]{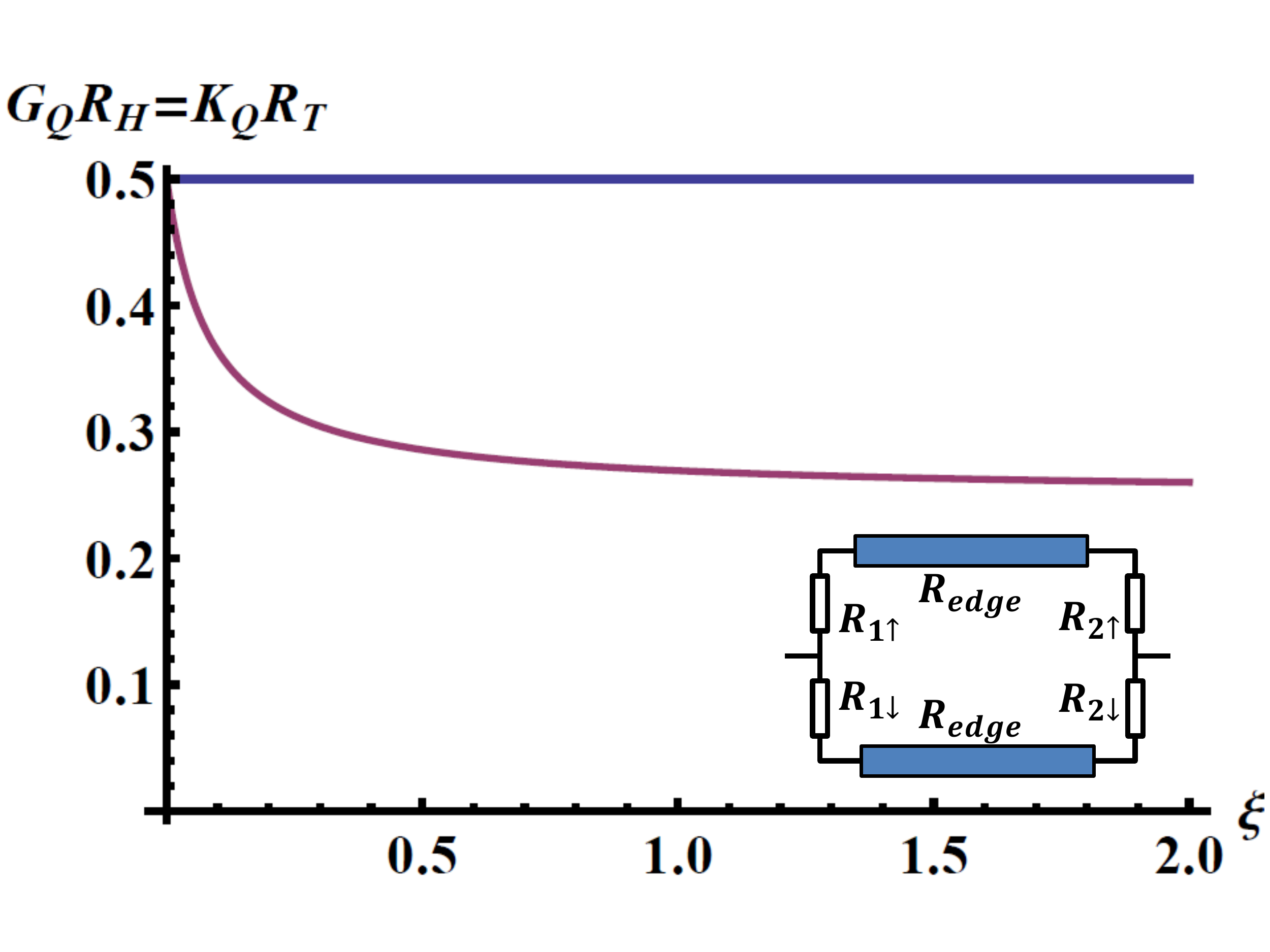}%
\vskip -0.4cm
\caption{(Color online) Dimensionless electrical and thermal Hall resistances as function of the spin scattering strength. The upper line corresponds to symmetric configuration with the polarizations of contacts $p_1=p_2=1$, where the Hall coefficients are independent of disorder. The lower curve corresponds to one electrode being a perfect ferromagnet and the other one being a nonmagnetic metal ($p_2=0$).   In that case, the Hall coefficients decrease with spin-scattering strength and saturate at strong spin scattering.   The total dimensionless conductances of the contacts  $g_{1}=g_{2}=1$, $\theta=0$ for all ferromagnets. Inset:  the effective electric circuit for large spin-scattering strength.  
\label{fig-RH_xi}}
\end{figure}
\begin{figure}
\includegraphics[width=8cm]{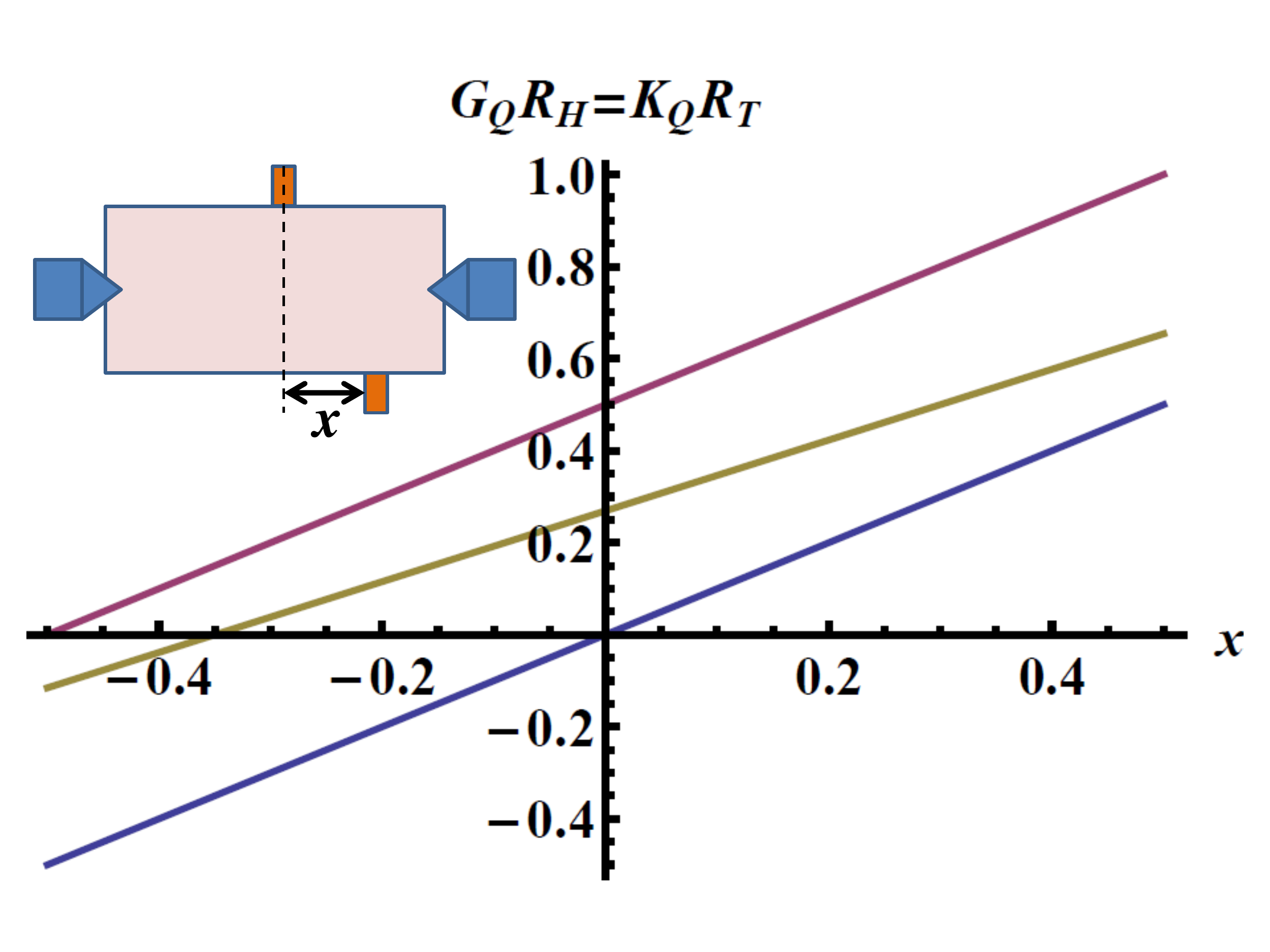}%
\vskip -0.4cm
\caption{(Color online) Dimensionless electrical and thermal Hall resistances as function of the position $x$ of the measurement contact on the lower edge (see Inset for the scheme of experimental setup).  The lower line describes the nonmagnetic electrodes ($p_1=p_2=0$), the upper line describes ideal FM electrodes with $p_1=p_2=1$, $\theta_1=\theta_2=0$, and the intermediate line corresponds to the left electrode being an ideal FM ($p_1=1$), and the right one being a nonmagnetic metal ($p_2=0$). The spin scattering strength $\xi=2$, the total dimensionless conductances of the contacts  $g_{1}=g_{2}=1$, and the length $L=1$ for all lines.   
\label{fig-RH_x}}
\end{figure}

\begin{figure}
\includegraphics[width=10cm]{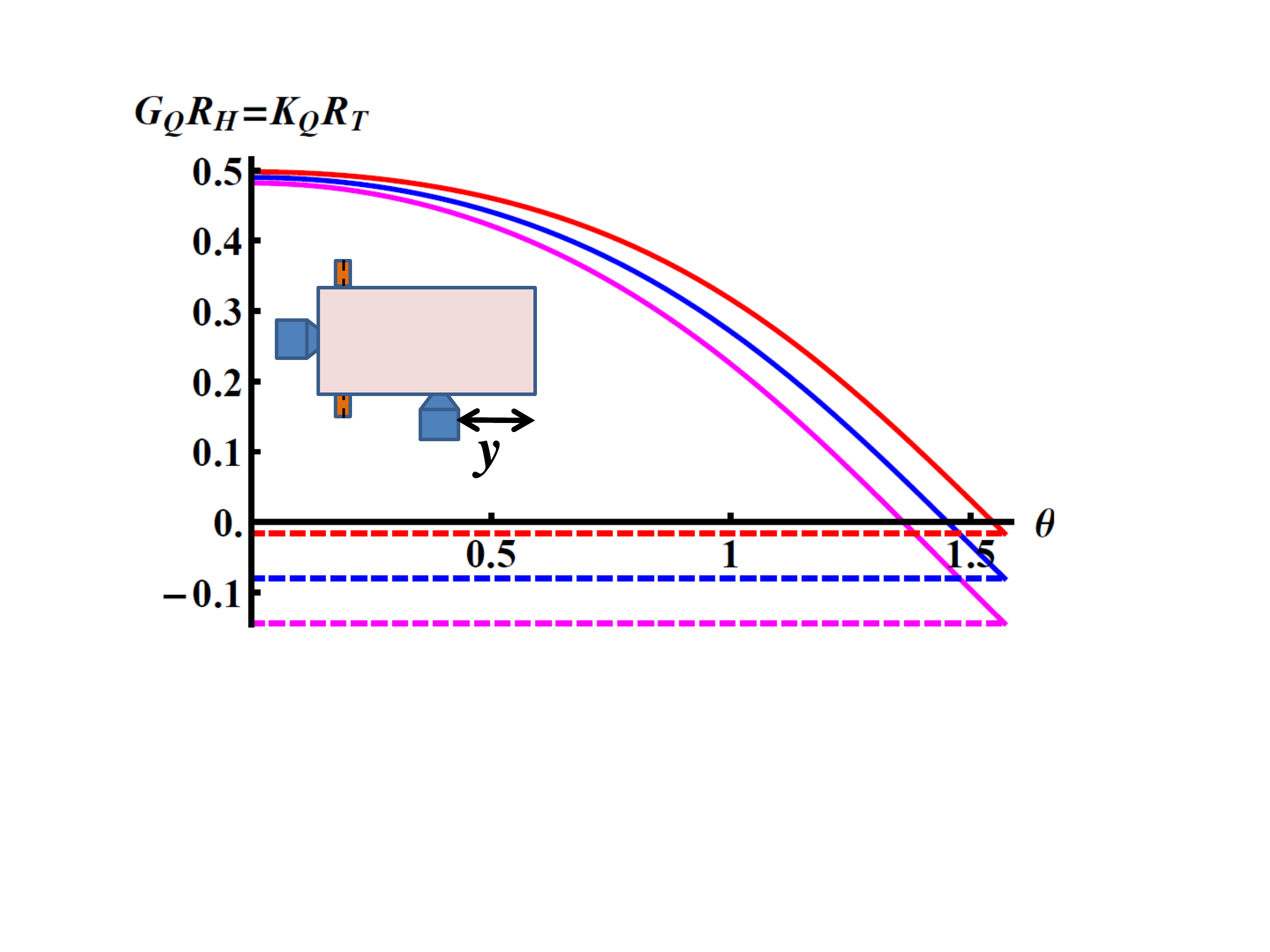}%
\vskip -2.0cm
\caption{(Color online) Dimensionless electrical and thermal Hall resistances as function of the angle $\theta$ for perfectly ferromagnetic ($p_1=p_2=1$) electrodes. 
The second lead is attached to  the lower edge of the sample at the distance $y$ to the right end (see the Inset, cf. Ref. \cite{Knez}).  $y=0.1$ for the upper curve, $y=0.5$ for the mid curve, $y=0.9$ for the lower curve. The dashed lines are the guidelines for the eye, indicating the result for the nonmagnetic electrodes at the corresponding values of $y$.  The spin scattering strength $\xi=1$, the total dimensionless conductances of the contacts  $g_{1}=g_{2}=1$, and the length of the sample $L=1$  for all curves.   
\label{fig-RH_theta-as}}
\end{figure}

To calculate the thermal and electrical  Hall resistances, we need expressions for the heat flow and the electrical current through the system. In the stationary state, the heat flow and the electrical current can be related to the particle flow at one of the contacts. For example, the particle flow out of the contact 1 (the left contact in Figs. \ref{fig-Setup}, \ref{fig-Scatt}) consists of spin-up electrons going from the contact to the lower edge, and spin-down electrons going  from the contact to the upper edge. Those particles have the distribution function of FM1. The flow into the contact consists of spin-up electrons coming from the upper edge and  spin-down electrons coming  from the lower edge, which are distributed according to the distribution functions $f_{u\uparrow}$ and $f_{d\downarrow}$ respectively. Therefore, the total electrical current and the heat flow through the contact are given by 
\begin{eqnarray*}
&& 
 I=\frac{e}{h}\int d\epsilon \left\{\left(g_{1\uparrow} + g_{1\downarrow}\right) F_1(\epsilon)-g_{1\uparrow} f_{u\uparrow}(\epsilon) -g_{1\downarrow}f_{d\downarrow}(\epsilon)\right\}, \label{I}\\ 
&&
 Q=\int \epsilon d\epsilon \left\{\left(g_{1\uparrow} + g_{1\downarrow}\right) F_1(\epsilon)-g_{1\uparrow} f_{u\uparrow}(\epsilon) -g_{1\downarrow}f_{d\downarrow}(\epsilon)\right\}. \label{Q}
\end{eqnarray*} 
Using Eq. (\ref{sol}), we obtain 
\begin{eqnarray}  
&&
I=\frac{e}{h}\left[g_{1\uparrow}\beta_{\mathrm{u}\uparrow}(-L/2) +g_{1\downarrow}\beta_{\mathrm{d}\downarrow}(-L/2)\right] (\mu_1-\mu_2), 
 \label{sol_I} \\
 \nonumber &&
Q=\frac{\pi^2 k_{\mathrm{B}}}{6 h}\left[g_{1\uparrow}\beta_{\mathrm{u}\uparrow}(-L/2)  +g_{1\downarrow}\beta_{\mathrm{d}\downarrow}(-L/2)  \right] (T^2_1-T^2_2), \\
 \label{sol_Q}
\end{eqnarray}
Calculating the relations between the transverse temperature gradient and the longitudinal heat current $R_T=\Delta T_{\perp}/Q$ and also between the Hall voltage and the electrical current   $R_{\mathrm{H}}=V_{H}/I$, we obtain the dependence of the Hall coefficients on the parameters of the experimental setup. Thereby the dimensionless thermal and electrical Hall resistances turn out to be equal to each other, as stated in Eq. (\ref{F}) with 
\begin{equation}
 \mathcal{F}=\frac{(\alpha_{\mathrm{d}\uparrow}(x)+\alpha_{\mathrm{d}\downarrow}(x)) -(\alpha_{\mathrm{u}\uparrow}(x)+\alpha_{\mathrm{u}\downarrow}(x))}{2 \left[g_{1\uparrow}\beta_{\mathrm{u}\uparrow}(-L/2)+g_{1\downarrow}\beta_{\mathrm{d}\downarrow}(-L/2)\right]}. 
\label{solF}
\end{equation}
Remarkably, the linear coordinate dependence of effective chemical potentials and effective temperatures (see Inset to  
Fig. \ref{fig-RH_theta}) cancels out exactly from the combination of coefficients in the numerator of Eq. (\ref{solF}), which results in the coordinate-independent Hall resistances.  This result reproduces the one obtained earlier by the zero-dimensional model of FM-TI-FM junction, where the coordinate dependence of distribution functions was neglected \cite{JMMM}. Moreover, as it is pointed out in Eq. (\ref{Fsym}),  the Hall resistances become independent of the relaxation time $\tau$ for the symmetric coupling between the FMs and TI, when the spin-polarizations and the angles between the magnetizations in both FMs and the spin-quantization axis in TI are equal. In contrast, for asymmetrical coupling between the FM-leads and TI, the Hall resistance is suppressed by spin-scattering. To describe this effect, we introduce the dimensionless measure of the spin-scattering strength as $\xi=L/(2 v\tau)$. The two lower curves in Fig. \ref{fig-RH_theta} show the angular dependence of the Hall resistances for two different strengths of spin-scattering in the case, when one electrode is taken to be a nonmagnetic metal. The maximal Hall resistance decreases for larger values of $\xi$. Similar dependence is illustrated in Fig. \ref{fig-RH_xi}. The upper line shows the independence of $R_H$ and $R_T$ of $\xi$ for the symmetric case, whereas the Hall resistances diminish sharply with $\xi$ in the asymmetric case, but further saturate to a finite value (see the lower curve in Fig. \ref{fig-RH_xi}). 

If one of the contacts is a fully polarized ferromagnet with magnetization parallel to the spin-quantization axis in TI, it couples to only one of the two helical edge states.  The other edge state, which is completely decoupled from that contact, does not contribute to the current in the absence of relaxation, $\tau\rightarrow\infty$. At the same time, it has the same chemical potential (temperature) on both edges. Thus, according to Eqs. (\ref{mu_eff}),  (\ref{T_eff}), the effective Hall voltage (transverse temperature gradient) diminishes twice. Then the Hall coefficients become perfectly quantized $G_Q R_H=K_qR_T=1/2$  independently of the characteristics of the other contact (the point $\xi=0$ in Fig. \ref{fig-RH_xi}).  

At strong spin-scattering, the Hall coefficients saturate at a nonuniversal value that depends on contact conductances. In that regime, the chiral character of the edge states is completely lost, and the measurement of the Hall coefficients is described by the effective electric circuit shown in the Inset to Fig. \ref{fig-RH_xi}. The saturation sets on, when the resistance along the edge exceeds by far the contact resistances. The resulting  value of the Hall resistance is given by 
\begin{equation}
R_H=\frac{1}{2}\left[\left(R_{1\downarrow} - R_{1\uparrow}\right)\left(1-\frac{x}{L}\right)+\left(R_{2\uparrow} - R_{2\downarrow}\right)\frac{x}{L}\right],  
\label{R_saturated}
\end{equation}
where $R_{i\sigma}$ denote the spin-resolved contact resistances. 
  
The results described above correspond to the symmetric position of the FM leads and measurement contacts situated at equal distance from the left edge of the sample (see Inset in Fig. \ref{fig-Setup}). In that case, the calculated transverse resistances coincide with the quantum Hall coefficients. Motivated by recent experiments  \cite{Knez}, we consider the experimental setups with asymmetric positions of the FM leads, and measurement contacts on the upper the lower edges  placed at different distances from the left end of the sample. 
In the latter case (see Inset to Fig. \ref{fig-RH_x}),  an additional potential difference is accumulated along the edge, which is proportional to the shift of the contacts with respect to each other. The transverse resistance is calculated according to Eq. (\ref{solF}), where the coefficients $\alpha_{\nu\sigma}(x)$ are taken at the coordinates of the corresponding measurement contacts, i.e.   $\alpha_{\mathrm{u}\sigma}(x_1)$, and  $\alpha_{\mathrm{d}\sigma}(x_2)$. The resulting transverse resistance depends linearly on the shift, as it is shown in Fig. \ref{fig-RH_x}. The upper and the lower lines show the cases of both leads being ideal ferromagnets or nonmagnetic metals respectively. They have equal slopes, which generalizes for all cases of symmetric electrodes. The intermediate line is obtained for an ideal FM as the left lead and a nonmagnetic metal as the right lead. It interpolates between the two above mentioned symmetric cases as  the measurement contact on the lower edge  approaches either the ferromagnetic 
or the nonmagnetic lead. 

Finally, we address the experimental geometry with asymmetric positions of the electrodes studied in Ref. \cite{Knez} (see Inset in Fig. \ref{fig-RH_theta-as}). In that case, the boundary conditions for the distributions functions at the shifted (the right) electrode  should be formulated at the coordinates, corresponding to the length of the path between the electrodes along the edge, that is $x=L/2-y$ for the functions $f_{\mathrm{d}\sigma}$, and $x=L/2+y$ for the functions $f_{\mathrm{u}\sigma}$ in the last two equations of Eqs. (\ref{BC}). Similarly to the case of the asymmetric position of the measurement contacts, shifting one of the electrode results in the appearance of different voltage drops along the opposite edges.  The detailed angular dependencies of Hall coefficients change at different values of $y$ retaining the same qualitative shape (see Fig. \ref{fig-RH_theta-as}). For the nonmagnetic electrodes, the results of Ref. \cite{Knez} can be reproduced by fitting the strength of the spin-scattering.

In conclusion, we worked out the theoretical description for a class of experimental setups that measure thermal and electrical quantum Hall coefficients in FM-TI-FM junctions. We predicted the values of Hall coefficients for various placements of the electrodes and measurement contacts, as well as for different magnetization directions and polarizations of FM leads. 
We showed theoretically that by changing directions of the magnetizations in FM  with respect to the spin-quantization axis in TI, one can obtain a large degree of  control over the generated Hall voltage and transverse temperature gradient. Of special importance for experimental measurements is the symmetric configuration of FM leads, in which case the Hall resistances are described by a universal function of magnetization direction, which is independent of spin-scattering strength. In the absence of spin-scattering,  one observes the quantized  values of electrical and thermal Hall coefficients, if one of the leads is a completely spin-polarized FM with magnetization direction parallel to the spin quantization axis in TI, independently of the properties of the other lead.  The proposed experimental setups lie well within the reach of modern technology.  
Considering different positions of the FM-leads and measurement contacts, we  reproduced results of experimental measurements with nonmagnetic leads attached asymmetrically to TI \cite{Knez}, and extended those results to the case of FM leads. Moreover, we also showed the appearance of nonzero Hall coefficients for asymmetric placement of measurement contacts along the opposite edges of the TI sample. This work gives impetus to the experimental realization of FM-TI-FM devices and their application in spintronics.

\begin{acknowledgments}
Authors acknowledge support from DFG through the Priority Program 1285 ``Semiconductor Spintronics'', and from SCE internal grant.  
\end{acknowledgments}

\end{document}